\documentclass[aps,prl,showpacs,twocolumn,superscriptaddress, amsmath, amssymb]{revtex4}

\usepackage{graphicx}
\usepackage{dcolumn}
\usepackage{bm}
\usepackage{epsfig}

\begin{document}

\preprint{}

\title{Stable nontrivial $Z_2$ topology in ultrathin Bi (111) films: a first-principles study}

\author{Zheng Liu}
\affiliation{State Key Laboratory of Low Dimensional
Quantum Physics, Tsinghua University, Beijing 100084, China}
\affiliation{Institute for Advanced Study, Tsinghua University,
Beijing 10084, China}
\author{Chao-Xing Liu}
\affiliation{Physikalisches Institut, Universit\"{a}t W\"{u}rzburg, D-97074 W\"{u}rzburg, Germany}
\author{Yong-Shi Wu}
\affiliation{Department of Physics and Astronomy,
University of Utah, Salt Lake City, 84112}
\affiliation{Department of Physics, Fudan University,
Shanghai 200433, China}
\author{Wen-Hui Duan}
\affiliation{State Key Laboratory of Low Dimensional
Quantum Physics, Tsinghua University, Beijing 100084, China}
\affiliation{Department of Physics, Tsinghua University, Beijing 10084,
China}
\author{Feng Liu}
\email{fliu@eng.utah.edu}
\affiliation{ Department of Materials Science and Engineering,
University of Utah, Salt Lake City, 84112 }
\author{Jian Wu}
\email{wu@phys.tsinghua.edu.cn}
\affiliation{State Key Laboratory of Low Dimensional
Quantum Physics, Tsinghua University, Beijing 100084, China}
\affiliation{Department of Physics, Tsinghua University, Beijing 10084,
China}

\date{\today}

\begin{abstract}
 Recently, there have been intense efforts in searching for new topological insulator (TI) materials. Based on first-principles calculations, we find that all the ultrathin Bi (111) films are characterized by a nontrivial $Z_2$ number independent of the film thickness, without the odd-even oscillation of topological triviality as commonly perceived. The stable nontrivial $Z_2$ topology is retained by the concurrent band gap inversions at multiple time-reversal-invariant k-points and associated with the intermediate inter-bilayer coupling of the multi-bilayer Bi film. Our calculations further indicate that the presence of metallic surface states in thick Bi(111) films can be effectively removed by surface adsorption.
\end{abstract}

\pacs{73.21.Ac, 71.70.-d, 73.20.At}
\maketitle

As a new insulating phase in condensed matter, the topological insulator (TI) has recently attracted a great deal of attention\cite{review}. The TI is distinguished from the conventional insulator by its unique gapless surface states residing in the middle of band gap as a consequence of the so-called $Z_2$ topology encoded in the wavefunctions.
There has been an intensive search for TI materials. Although quite a few compounds have been found to be 3D TIs\cite{Fu07b, Hsieh08, Hsieh09, zhang09, Chen09}, up to now only the HgTe quantum well is verified to be a 2D TI (or a quantum spin Hall insulator) experimentally\cite{Konig07}. Recently, Murakami predicted a single bilayer (BL) Bi (111) film to be an elemental 2D TI \cite{Murakami06, Murakami11}, and further speculated the multi-BL Bi (111) film to exhibit an odd/even oscillation of topological triviality with film thickness\cite{Murakami06} by considering the multi-BL film as a stack of BLs with no or very weak inter-BL coupling \cite{Fu07a}. Generally speaking, the 2D TI phase in thin films is commonly perceived to depend sensitively on film thickness, as shown for the $Bi_2Se_3$ and $Bi_2Te_3$ ultrathin films \cite{Liu10}. The requirement of fine tuning the thickness makes the experimental fabrication of 2D TIs rather difficult. Therefore, it is desirable to search for new materials or new schemes to obtain a 2D TI.

In this Letter, based on the first-principles calculations of band structures and wavefunction parities, we find that in fact all the Bi (111) ultrathin films are characterized by a nontrivial $Z_2$ number independent of the film thickness. The films with 1- to 4-BL are intrinsic 2D TIs, while those with 5- to 8-BL are 2D TIs sandwiched with trivial metallic surfaces that can be extrinsically removed by surface adsorption. This finding is in direct contrast to the odd/even oscillation of topological triviality speculated for the Bi films \cite{Murakami06}, as well as to the thickness-dependent topology shown for other 2D films \cite{Liu10}. The surprisingly stable 2D TI phase in the Bi(111) films are found to be retained by a concurrent band gap inversion at multiple time-reversal-invariant k-points (TRIKs) when the film thickness is increased. A detailed analysis of the 2- and 3-BL Bi films indicates that the intermediate inter-BL coupling plays an important role in defining their unique topological property.

Bismuth is one of the main group elements that has the strongest spin-orbit coupling (SOC), a fundamental mechanism to induce the $Z_2$ topology. For this reason, many 3D TIs make use of Bi, even though 3D bulk Bi itself is topologically trivial \cite{Fu07b}. The electronic properties of Bi, such as the bulk band structure \cite{Biband}, the surface states \cite{Surface} and the semimetal-to-semiconductor transition \cite{SMSC, Blugel08}, have been well established in the literature. The outermost shell of Bi has the electron configuration $6s^26p^3$. It tends to form three bonds to close the shell. The single-crystal Bi has a bi-layered structure, with an ABC stacking sequence along the (111) direction (Fig. 1a). Within each BL, every Bi atom forms three $\sigma$  bonds with its nearest neighbors in a trigonal pyramidal geometry. Projecting onto the (111) plane, the BL forms a hexagonal lattice with two atoms per unit cell (Fig. 1b). There are three key structural parameters to define the lattice: the in-plane lattice constant $a$, the intra-BL bond angle $\alpha$ (or the intra-BL height difference $d_1$), and the inter-BL spacing $d_2$. Our calculated Bi crystal structural parameters as shown in Fig. 1a and b agree well with previous calculations \cite{Surface}. The ultra-thin Bi (111) thin films, consisting of a few number of stacked BLs, have slightly relaxed structural parameters relative to the bulk values. Specifically, $\alpha$ approaches $90^o$ and $d_2$ increases by about $6\%$. This implies a slightly enhanced p-orbital feature of the intra-BL bonds and a weakening of the inter-BL coupling.

\begin{figure}
\includegraphics[width=0.4\textwidth]{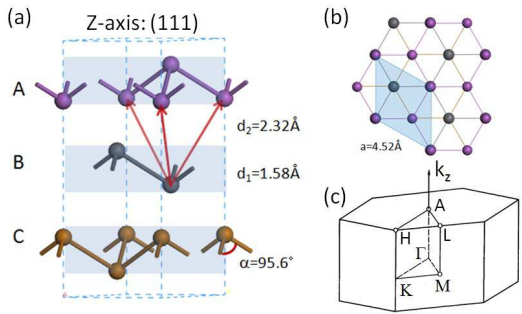}
\caption{\label{fig:unitcell} (a) The hexagonal unit cell of single-crystal bismuth. (b) The top view of the Bi lattice. (c) The first Brillouin zone of the hexagonal lattice. }
\end{figure}

To identify the 2D TI phase, a single $Z_2$ topological number ($\nu$) is used as the ``order parmeter'' \cite{Fu07a}: $\nu=1$ indicates a topologically nontrivial phase; $\nu=0$ indicates a trivial phase. The calculation of the $Z_2$ number can be dramatically simplified by the so-called ``parity method'' \cite{Fu07b}, if the system is space inversion invariant, as the case of the Bi (111) film. Accordingly, the $Z_2$ number of Bi films can be obtained from the wavefunction parities at four TRIKs ($K_i$), one $\Gamma$ and three $M$s, as

\begin{eqnarray}
\delta(K_i)&=&\prod_{m=1}^N\xi_{2m}^i\nonumber\\
(-1)^\nu&=&\prod_{i=1}^4\delta(K_i)=\delta(\Gamma)\delta(M)^3\nonumber
\end{eqnarray}
where $\xi=\pm$ is the parity eigenvalues and $N$ is the number of the occupied bands.

Single-electron wavefunction parities are calculated within the density functional theory using the plane wave basis, as implemented in the ABINIT package \cite{abinit}. We employ the local density approximation (LDA) \cite{LDA} and the Hartwigsen-Goedecker-Hutter pseudopotential \cite{HGH}, which is generated on the basis of a fully relativistic all-electron calculation and tested to be accurate for heavy elements like Bi. The spin-orbit coupling is included in the self-consistent calculations as described in \cite{soc}. To model the thin film, a supercell of slab is used with periodic boundary conditions in all three dimensions with a $10~\AA$ thick vacuum layer in the (111) direction to eliminate the inter-slab interaction. A plane wave cutoff of $24~Ry$ and a $\Gamma$-centered k-point mesh of $10\times10\times1$ are used. All the atomic positions are fully relaxed for each film.

\begin{table}
\caption{\label{count1} The total parity at the $\Gamma$ and $M$ points and the $Z_2$ number of Bi (111) films with different thickness}
\begin{ruledtabular}
\begin{tabular}{ccccccccc}
  \# of BLs & 1 & 2 & 3 & 4 & 5 & 6 & 7 & 8\\ \hline
  $\delta$($\Gamma$) & + & + & - & - & + & + & - & -\\
  $3\delta$($M$)      & - & - & + & + & - & - & + & +\\
  $\nu$              & 1 & 1 & 1 & 1 & 1 & 1 & 1 & 1
\end{tabular}
\end{ruledtabular}
\end{table}

Table 1 shows the calculated $Z_2$ numbers for 1- to 8-BL films.  Surprisingly, all the films we calculated are characterized by the nontrivial $Z_2$ number ($\nu=1$), in direct contrast to the oscillation of topological triviality as commonly perceived. We notice that the total parity $\delta(\Gamma)$ and $\delta(M)$ simultaneously change their signs for every 2 BLs. This kind of parity oscillation, which originates from the inverted band gap of a 3D TI, has been reported in $Bi_2Se_3$ and $Bi_2Te_3$ ultrathin films, but only at the $\Gamma$ point \cite{Liu10}. The uniqueness of Bi film is the inverted band gaps both at the $\Gamma$ and $M$ points due to the strong SOC of Bi \cite{Fu07b}. As a consequence, the parity oscillation under the quantum confinement occurs concurrently at all TRIKs (one $\Gamma$ and three Ms) as the film thickness increases. Being the product of $\delta(\Gamma)$ and $\delta(M)$, the $Z_2$ number shows up as the ``beat" of two oscillations, which makes the 2D TI phase in Bi ultrathin films much more stable than in the $Bi_2Se_3$ and $Bi_2Te_3$ films. We expect that the $Z_2$ number of Bi films will eventually change at some point when the phase difference between the oscillations at $\Gamma$ and $M$ accumulates to $\pi$, but this requires calculations of much thicker Bi films possibly beyond the current computational capability.

\begin{figure}
\includegraphics[width=0.4\textwidth]{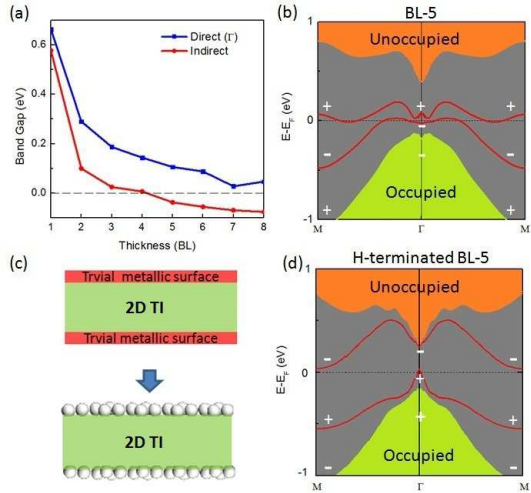}
\caption{\label{fig:2} (a)The direct ($\Gamma$-point) and indirect band gap as a function of the film thickness. (b) The band structure and the parity information of 5-BL. (c) A schematic depiction of the electronic property of an Bi (111) film before and after the H-termination. (d) The band structure and the parity information of H-terminated 5-BL. }
\end{figure}

In Fig. 2a, we plot both the direct and indirect band gaps as a function of the film thickness. Within the calculated film thickness range, the direct gap always remains open, which is essential for a well-defined $Z_2$ number. Below 4 BLs, the film is a semiconductor having a nontrivial $Z_2$ number, and hence representing an intrinsic 2D TI. However, the indirect band gap becomes negative above 4 BLs, leading to a semiconductor-semimetal transition\cite{Blugel08}. The semi-metallization arises from two overlapping bands around the Fermi level, as shown in Fig. 2b for the 5-BL film as an example. It has been pointed out \cite{Surface, Blugel08} that these two bands have evident surface band features and tend to become gapless at the $\Gamma$ and M points in the limit of a semi-infinite system. From this view, the films from 5- to 8-BL can be regarded as a 2D TI sandwiched between two trivial metallic surfaces (top panel, Fig. 2c). The meaning of ``trivial'' here is two-fold. On the one hand, if we consider the surface as an individual 2D system, its $Z_2$ number is 0 (trivial), as obtained from the surface band parities at the $\Gamma$ and M points (see Fig. 2b). Therefore, the surface bands have no contribution to the nontrivial $Z_2$ number of the film. On the other hand, in the limit of semi-infinite system, because bulk Bi is a 3D $Z_2$ topologically trivial insulator, these trivial metallic surface bands are not robust as those of a 3D strong TI and hence, in principle, can be easily removed by surface defects or impurities, e.g. via surface adsorption. To test this idea, we have terminated the two surfaces with H atoms as schematically depicted in the bottom panel of Fig. 2c and repeated the calculation. We find that upon surface adsorption, the two surface bands are separated apart opening a gap around the Fermi level, as shown in Fig. 2d. Meanwhile, the H atoms introduce additional occupied bulk bands, which are found to be topologically trivial, so that the $Z_2$ number remains nontrivial (see Fig. 2d). Thus, by the extrinsic effects of H surface adsorption, the thicker films above 5 BLs are effectively converted into 2D TIs similar to those ultrathin films below 4 BLs. We note that because LDA is known to underestimate the band gap, the actual transition thickness is likely to be thicker than the 4-to-5 BLs, but the overall trend of behavior we show here should remain valid.

\begin{figure}
\includegraphics[width=0.4\textwidth]{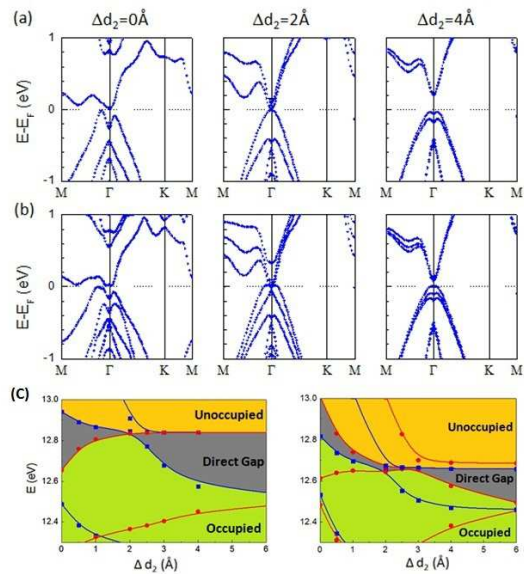}
\caption{\label{fig:3} The band structure of (a) 2-BL and (b) 3-BL under different inter-BL spacing. (c) The energy level at the $\Gamma$ point of (left) 2-BL and (right) 3-BL as a function of the inter-BL spacing. The red (circle) lines indicate even-parity levels. The blue (square) lines indicate odd-parity level. }
\end{figure}

It has been predicted that the 1-BL Bi (111) film is a 2D TI \cite{Murakami06, Murakami11}. If we could regard the n-BL film as a stack of these nontrivial 1-BL films and without inter-BL coupling, all the bands would have n-fold degeneracy and there would be naturally an odd-even oscillation of $Z_2$ topology: $\nu=1$ for the odd-BL stacks and $\nu=0$ for the even-BL stacks \cite{Fu07a} as every additional BL flips the $Z_2$ number. Such odd-even oscillation of $Z_2$ topology from the zero inter-BL coupling limit can be extended to the weak inter-BL coupling regime under the adiabatic approximation, which was speculated to be applicable to the Bi(111) films \cite{Murakami06}. However, our first-principles results show that the Bi (111) film may not be adiabatically connected to the zero inter-BL coupling limit.

The Bi (111) film actually represents a special interesting class of films having an intermediate inter-BL coupling strength. The inter-BL bond energy is calculated to be ~0.3-0.5 eV per bond, which is noticeably larger than the typical values of weak interfacial bonds, such as Van der Waals bond, but smaller than the values of typical chemical bonds. This intermediate inter-BL coupling may have a significant influence on the topological property. To reveal the influence, we have calculated a set of ``model'' Bi films with their inter-BL coupling tuned gradually from the real intermediate regime to the hypothetical weak coupling regime that could be adiabatically connected to the zero coupling limit. This is done by artificially increasing the inter-BL distance from the equilibrium value $d_2$ with an increment of $\Delta d_2$, using the supercell technique. Here we take 2-BL (3-BL) film as an example of even (odd) number of BL films. Figure 2a and 2b show the evolution of band structures as a function of $\Delta d_2$ for the 2-BL and 3-BL films, respectively. For $\Delta d_2 =0$, the figures show the realistic band structures of the 2-BL and 3-BL films, while for $\Delta d_2 = 4 \AA$, the corresponding band dispersion of hypothetical films shows a weak inter-BL coupling case that is adiabatically connected to the zero coupling limit, based on topological analysis. We note that as the $\Delta d_2$ is tuned from $4 \AA$ to $0 \AA$, the direct gap at $\Gamma$ point undergoes a closing and re-opening process, indicating that the realistic band dispersion of Bi films may not be adiabatically connected to the zero coupling limit.


To trace the change of topology with the inter-BL coupling, we analyze the parity property of the energy levels at $\Gamma$ point as a function $\Delta d_2$ in Fig. 2c for the 2-BL (left) and 3-BL (right) Bi films, respectively. When $\Delta d_2$ is reduced, the n-fold degeneracy for the zero coupling limit as mentioned above is lifted by the inter-BL coupling. Although the variation of band gaps is similar for the 2-BL and 3-BL films, the change of total parity of all the occupied bands shows quite different behaviors. For the 2-BL film, each doubly degenerate level is split into one odd-parity sub-level with higher energy and one even-parity sub-level with lower energy. Consequently, the lowest unoccupied level and the highest occupied level have the opposite parities (left panel of Fig. 2c). As $\Delta d_2$ is reduced, a level crossing and hence a parity exchange happens at $\Delta d_2=2 \AA$ due to the opposite parities of the two crossing levels, leading to a change of $\nu$ from 0 (for the non-coupling even-BL films) to 1, and hence converting the $Z_2$ number of 2-BL film from trivial to nontrivial. The situation, however, is different for the 3-BL film. Now, each 3-fold degenerate level is split into three sub-levels, always with one odd-parity sub-level sandwiched by two even-parity sub-levels (right panel of Fig. 2c). Consequently, both the lowest unoccupied and highest occupied levels have the same even parity. As $\Delta d_2$ is reduced, a level anti-crossing instead of crossing happens and the number of odd- and even-parity sub-levels for both the occupied and unoccupied bands remain unchanged, resulting in no change to the non-trivial $Z_2$ topology of the 3-BL film.

In the above simple picture, we have implicitly used the prerequisite condition that the level crossing or anti-crossing happens only at the $\Gamma$ point but not at the M point and involves only a few levels close to the band gap, which can only be satisfied by an inter-BL coupling that is not too strong. Also, the coupling can not be too weak in order to move away from the weak coupling regime that is adiabatically connected to the zero coupling limit. Therefore, the intermediate coupling strength is a mandatory condition for the non-trivial $Z_2$ topology. Such "intermediate coupling principle" may be utilized in search for the 2D TI phase in other materials.

As a summary, the film below 4 BLs is an intrinsic 2D TI with the band structure consisting of "molecular orbital" levels without distinction of surface bands from bulk bands, as shown in Fig. 3a (left panel) and 3b (left panel) for the 2-BL and 3-BL film, respectively. Above 4 BLs, the band structure is made of surface bands superimposed onto a 2D projected bulk band, as shown in Fig. 2a for the 5-BL film. The projected 2D bulk bands keep the non-trivial topology of a 2D TI with a sizable gap, but the surface bands gradually appear in the middle of the projected bulk band gap with the increasing film thickness (see also \cite{Blugel08}), leading to a semiconductor (1- to 4-BL) to a semimetal (5- to 8-BL) transition. The trivial metallic surface states can be removed by surface H adsorption (Fig. 2d), which effectively converts the Bi films into true 2D TIs.


Our finding of all the ultrathin Bi (111) films being 2D TIs independent of thickness may provide a possible explanation of the recent observation of 1D topological metal on the Bi (114) surface\cite{exp114}, and will stimulate more experimental interest in this intriguing system. The Bi (111) films above 6-BL have already been successfully grown via molecular beam epitaxy \cite{exp112}, and hopefully even thinner Bi films can be grown in the near future. The physical mechanism we identified for retaining the stable nontrivial $Z_2$ topology has broad implications. Most importantly, tuning the inter-layer coupling to the intermediate regime, so as to remove the odd-even oscillation of topological triviality, can be applied as a general strategy to obtain the TI phase.

We thank Xiaofeng Jin for stimulating discussions and for sharing with us their unpublished experimental data on the Bi thin films. The work in Tsinghua University is supported by the Ministry of Science and Technology of China (Grants No. 2011CB606405, 2011CB921901, and 2009CB929401) and NSF of China (Grants No. 10974110 and 11074139). CXL acknowledges the support from Humboldt foundation, YSW is supported in part by US NSF grant PHY-0756958, and FL acknowledges supports from the DOE-BES program.

\end{document}